\documentstyle[12pt]{article}
\begin{document}
\vsize=25.0 true cm
\hsize=16.0 true cm
\predisplaypenalty=0
\abovedisplayskip=3mm plus 6pt minus 4pt
\belowdisplayskip=3mm plus 6pt minus 4pt
\abovedisplayshortskip=0mm plus 6pt
\belowdisplayshortskip=2mm plus 6pt minus 4pt
\normalbaselineskip=14pt
\normalbaselines
\begin{titlepage}
 
\begin{flushright}
{\bf CERN-TH/99-120 \\
IFT/99-06}
\end{flushright}
 
\vspace{0.5cm}
\begin{center}
{\bf\Large
Library of SM and anomalous $WW\gamma$ couplings 
for the $e^+e^- \to f \bar f n\gamma$ Monte Carlo programs
}\end{center}
 \vspace{0.8cm}
\begin{center}
  {\bf A. Jacholkowska} \\
{\em Laboratoire de l'Acc\'el\'erateur Lineaire, 
CNRS-IN2P3, 914050 Orsay, France}\\
\vspace{0.2cm}
   {\bf J. Kalinowski} \\
{\em Instytut Fizyki Teoretycznej, Ho\.za 69, 00681 Warszawa, Poland}\\
and\\
\vspace{0.2cm}
   {\bf Z. W\c{a}s } \\
  {\em CERN, Theory Division, CH 1211, Geneva 23, Switzerland,\\ and}\\
   {\em Institute of Nuclear Physics,
        Krak\'ow, ul. Kawiory 26a, Poland}\\
\end{center}
 
\vspace{1.5cm}
\begin{center}
{\bf   ABSTRACT}
\end{center}
A brief description of the library of the Standard Model and anomalous
$WW\gamma$ coupling contribution to the matrix element for 
$e^+e^- \to \nu\bar{\nu} n\gamma$ process is given. It can be
used with any  Monte Carlo program for $e^+e^- \to f \bar f n\gamma$ 
processes. A working example of the application for the KORALZ version
4.04 is also provided.

\vskip 2.3 cm
\centerline{\it To be submitted to Computer  Physics Communications}
 \vspace{1.4cm}
\begin{flushleft}
{\bf 
 CERN-TH/99-120\\
IFT/99-06 \\
May 1999}
\end{flushleft}
\footnoterule
\noindent
{\footnotesize
\begin{itemize}
\item[${\dagger}$]
Work supported in part by 
Polish Government grants 
KBN 2P03B08414, 
KBN 2P03B14715, 
KBN 2P03B03014 (JK), 
Maria Sk\l{}odowska-Curie Joint Fund II PAA/DOE-97-316,
and Polish-French Collaboration within IN2P3.
\end{itemize}
}
 
\end{titlepage}

 \vskip 10pt
\section{Introduction}

The $\nu \bar \nu \gamma $ production in $e^+e^-$ collisions is of
great interest, as it is sensitive to the triple gauge boson coupling
$WW\gamma$ of the Standard Model. Its precise measurement will not serve
 only as a stringent test of the SM, but may reveal (or constrain)
anomalous gauge couplings. The events with photon(s) plus missing
energy might originate also from other mechanisms, signalling new
physics beyond the Standard Model. For example, such final states can
be produced in both gravity- and gauge-mediated supersymmetric models
or low-scale gravity. The missing energy in these events is caused by
weakly interacting supersymmetric particles, such as gravitinos,
neutralinos and/or sneutrinos \cite{others}, or gravitons
\cite{lowsg}. In all such cases the Standard Model $e^+e^- \to \nu
\bar \nu \gamma$ events are irreducible background and reliable
theoretical predictions for them are therefore necessary.

With the sensitivity afforded by the LEP
experiments~\cite{gam183,refWW}, and expected at future $e^+e^-$ 
colliders~\cite{acco}, the photon(s) plus missing energy events
provide an opportunity to search for new physics phenomena. Any
meaningful interpretation of the experimental data requires a Monte
Carlo simulation in which Standard Model predictions may be augmented
by the contributions from possible anomalous couplings.

Since the LEP collaborations are entering their final years of
operation, now is a good time to document the programs that have
actually been used in the data analyses at LEP. In the present paper
we document the library for the calculation of the effects of the
$WW\gamma$ interaction for the $e^+e^- \to \nu \bar \nu n\gamma$ process
within the Standard Model as well as from the
anomalous couplings.
It is based on the work of \cite{nunug} and the description of the
physical content of the program interface and discussion of its
uncertainties can be found in \cite{Jachol}. For an alternative 
implementation of the $WW\gamma$ vertex, see \cite{mmnp}.

In principle our library can be combined with any $e^+e^- \to f \bar f
\ n\gamma$ generator, but in the present paper we will use an
interface to KORALZ version 4.04, described in~\cite{KORALZ,KORALZ1},
as a working example. That is the reason why the fortran code of the
library will be archived together with the KORALZ tree of subdirectories
~\cite{KORALZ1}. 
Let us note that, in future, KORALZ
will be replaced by a new program, KK2f~\cite{KK2f}, which is based on a
more powerful exponentiation at the spin amplitude level and in which the
library will also be easy to implement.

Version 4.04 of the KORALZ Monte Carlo program can be used to simulate
$e^+e^- \to f \bar f \ n\gamma$, ($f=\mu,\tau,u,d,c,s,b, \nu$)
processes up to the LEP2 energy range, including YFS exclusive
exponentiation of initial- and final-state bremsstrahlung and,
optionally, the effects of various anomalous couplings. In the case of the
LEP2 centre-of-mass energies and if $f=\nu$, the present library can be
used for that purpose.

\section{Calculation of anomalous couplings}

To evaluate the effects of anomalous $WW\gamma$, a tree-level
calculation of the squared matrix element for the process $e^+ e^-
\rightarrow \nu \bar \nu \gamma$ has been carried out~\cite{nunug}. It
includes the effects of the anomalous C- and P-conserving\footnote{For a
recent phenomenological analysis of CP-violating couplings in $e^+e^- \to \nu
\bar \nu \gamma$, see \cite{dkk}.} 
contributions parametrized with the help of the
couplings  $\Delta\kappa_\gamma$, $\lambda_\gamma$ (in what follows we
will suppress the subscript $\gamma$). The library is formed on this
basis. When activated, it uses the 4-momenta of the neutrinos and the
photon provided by the host program to compute a weight, $w$, for each
event according to
\begin{equation}
w = \frac{| {\cal M}_{\mathrm SM}^{WW\gamma \ excl.} 
+ {\cal M}_{\mathrm SM}^{WW\gamma }+{\cal M}_{\mathrm ano}^{WW\gamma }
 | ^2 } {| {\cal M}_{\mathrm SM}^{WW\gamma \  excl.} | ^2}.
\end{equation}
${\cal M}_{\mathrm ano}^{WW\gamma }$ is the matrix element due to the
anomalous $\Delta\kappa \neq 0$, $\lambda \neq 0$ couplings, the
${\cal M}_{\mathrm SM}^{WW\gamma }$ is the matrix element due to the
SM $WW\gamma$ interaction and ${\cal M}_{\mathrm SM}^{WW\gamma \
excl.}$ represents the remaining part of the SM matrix element for the
$e^+e^- \to \nu \bar \nu \gamma$ process\footnote{Note that such a
separation is gauge-dependent and, if not treated carefully, could
lead to meaningless results. See ref. \cite{Jachol} for details.}.
Note that because of the above separation, even for the Standard Model
$WW\gamma$ interaction, the use of the library is necessary to
calculate the weight $w$.

As the calculation of ref. \cite{nunug} is performed at ${\cal O}
(\alpha)$, the case of multiple bremsstrahlung requires a special
treatment. In this case, a reduction procedure is first 
applied in
which all photons except the one with the highest-$p_T$ are
incorporated into the 4-momenta of effective beams.  In the second
step, the 4-momenta of
the highest $p_T$ photon, the effective beams and neutrinos are then
used to compute the weight. Cross-checks of the calculation as well as
checks of the validity of the reduction procedure are described
in~\cite{Jachol}. The results of the calculation have been used in the
measurement of the $WW\gamma$ coupling described in~\cite{gam183}.

\section{Flags to control anomalous couplings in KORALZ}
The calculation of weights for anomalous couplings is activated by setting
the card {\tt IFKALIN=1}. This is transmitted from the main program
via the KORALZ input parameter {\tt NPAR(15)}. Additional input
parameters are set in the routine {\tt kzphynew(XPAR,NPAR)}, although
there are currently no connections to the KORALZ matrix input
parameters {\tt XPAR} and {\tt NPAR}~\footnote{ In  most uses
of KORALZ, the numerical value of these parameters is irrelevant or 
defaults are sufficient. It is expected that the advanced user may like
to change them, connecting directly the {\tt kzphynew(XPAR,NPAR)} routine
with her or his main program. }.
 Table~\ref{tab:parameters} summarizes the functions  
of these input parameters.

\begin{table}[hbt!]
\setlength{\tabcolsep}{0.5mm}
\renewcommand{\arraystretch}{1.6}
\begin{center}
{

\begin{tabular}{|l| l| c| } \hline

Parameter &      Description        &   Default \\ \hline
IENRICH   & \parbox{0.7\textwidth} {Enrich spectrum of generated 
            sample with  hard-photon events IENRICH$ =$ 1/0 (on/off)} 
          &  0 \\ \hline
IRECSOFT  & Generate {\em only} events with photon(s) if IRECSOFT$ =$ 1 
          & 0\\ \hline
EMINACT   & \parbox{0.7\textwidth} {Minimum sum of all photon energies 
            required to calculate anomalous weights}             
          &  17 GeV   \\ \hline
EMAXACT   & \parbox{0.7\textwidth} {Maximum sum of all photon energies 
            allowed to calculate anomalous weights}             
          & 1000 GeV  \\ \hline
PTACT     & \parbox{0.7\textwidth} {Minimum sum of all photon momenta 
            transverse to the  beam direction required to calculate 
            anomalous weights} 
          & 2 GeV      \\ \hline
\end{tabular}
}

\caption{Input parameters to control the calculation of weights for anomalous 
$WW\gamma$ couplings.}

\label{tab:parameters}
\end{center}
\end{table}
More input parameters are initialized in the subroutine {\tt anomini}, 
which is placed in  the file {\tt gengface.f} 
and subroutine {\tt initialize} of the file {\tt geng.f}. Both files
are placed in  the directory 
{\tt korz{\_}new/nunulib}.

In order to provide the user with enough information to retrieve the $w$
for a given event for any $\Delta\kappa$, $\lambda$, we take 
advantage of the fact that, for each event, one may write the $w$ as a
quadratic function of the anomalous couplings, in terms of the
results calculated for six numerically distinct combinations of the
$\Delta\kappa$, $\lambda$ values as follows:
\begin{eqnarray} \label{eqn:weight}
w(\Delta\kappa,\lambda)&= & 
    \Bigl(1-\bigl({\lambda \over \lambda_0}\bigr)^2-\bigl({\Delta\kappa \over
     \Delta\kappa_0}\bigr)^2 +{\lambda \over
     \lambda_0}{\Delta\kappa \over
                    \Delta\kappa_0}\Bigr)  w( 0, 0)\nonumber
          - \Bigl({\Delta\kappa \over 2\Delta\kappa_0}-
                    {1 \over 2}\bigl({\Delta\kappa \over
                    \Delta\kappa_0}\bigr)^2\Bigr)  
                   w(-\Delta\kappa_0,0)\nonumber\\ 
            &+&\Bigl( {\Delta\kappa \over
                    2\Delta\kappa_0}+ {1 \over 2}\bigl({\Delta\kappa \over
                    \Delta\kappa_0}\bigr)^2- {\lambda \over
                    \lambda_0}{\Delta\kappa \over
                    \Delta\kappa_0}\Bigr)  w( \Delta\kappa_0, 0) 
           -  \Bigl({\lambda \over 2\lambda_0}-
                    {1 \over 2}\bigl({\lambda \over \lambda_0}\bigr)^2\Bigr) 
                    w( 0,-\lambda_0)\nonumber\\ 
           &+&\Bigl( {\lambda
                    \over 2\lambda_0}+ {1 \over 2}\bigl({\lambda \over
                    \lambda_0}\bigr)^2 -{\lambda \over
                    \lambda_0}{\Delta\kappa \over
                    \Delta\kappa_0}\Bigr)  w( 0, \lambda_0)  
                   +{\lambda \over
                    \lambda_0}{\Delta\kappa \over \Delta\kappa_0} 
                    w(\Delta\kappa_0, \lambda_0).
\end{eqnarray}
When the calculation is completed, 
the six weights  are stored in the common block
{\tt common /kalinout/ wtkal(6)}, with the assignments shown
in Table~\ref{tab:weights}.
\begin{table}[hbt!]
\begin{center}
\begin{tabular}{|l|l|} \hline

Common block entry   &  Weight parameter \\ \hline 
{\tt wtkal(1)}  &  $w(0,0)$                 \\ 
{\tt wtkal(2)}  &  $w(-\Delta\kappa_0,0)$         \\
{\tt wtakl(3)}  &  $w(\Delta\kappa_0,0)$        \\    
{\tt wtkal(4)}  &  $w(0,-\lambda_0)$          \\
{\tt wtkal(5)}  &  $w(0,\lambda_0)$         \\
{\tt wtkal(6)}  &  $w(\Delta\kappa_0,\lambda_0)$  \\  \hline

\end{tabular}
\caption{Correspondence between entries in {\tt kalinout} common block entries
and weight parameters of eq.~\ref{eqn:weight}.}

\label{tab:weights}
\end{center}
\end{table}
The user is then free to calculate the $w$ for whatever combination of
$\Delta\kappa$ and $\lambda$  is desired. In our program we set
$\Delta\kappa_0=10$ and $\lambda_0=10$. If the {\tt IENRICH} input
parameter is set to 1 the generated sample will have more events with
hard photons than predicted by the Standard Model. The appropriate
compensating factor is included into the weights {\tt wtkal}. It
is thus always assured that e.g. the generated sample, if the weight
{\tt wtkal(1)} is used, represents the Standard Model predictions.

The code for a calculation of the weight $w$ is placed in the directory {\tt
korz{\_}new/nunulib} in the files {\tt geng.f} and {\tt gengface.f}.
\\

\section{Demonstration programs}
The demonstration program {\tt DEMO3.f\/} for the run 
of KORALZ when our library is activated 
can be found in the directory 
{ \tt korz{\_}new/february} and the output {\tt DEMO3.out}
in the directory { \tt korz{\_}new/february/prod1}. The {\tt DEMO.f} for 
the run of KORALZ 
with the Standard Model interactions only
and its output {\tt DEMO.out} are also included in the directories
mentioned above. All these files as well as the library itself
are archived together with KORALZ \cite{KORALZ1}.

\vskip 1cm
\noindent  {\bf \large Acknowledgements}\\
One of us (ZW) thanks CERN Theory Division for support during the final 
work on the paper.


\end{document}